\documentclass[aps,showpacs,11pt]{revtex4}
\usepackage{dcolumn}
\usepackage{graphicx}
\usepackage{amsmath}
\usepackage{amsfonts}
\usepackage{amssymb}
\usepackage{psfrag}
\usepackage{wrapfig}
\usepackage{subfigure}
\usepackage{makeidx}
\usepackage{bm}
\usepackage{epsf}
\usepackage{hyperref}
\usepackage{color}
\usepackage{multirow}
\newcommand{\be}{\begin{equation}}
\newcommand{\ee}{\end{equation}}

\begin{document}

\begin{abstract}

We investigate the impact of scalar-tensor theories on the 21cm brightness temperature. We consider a specific model in Horndeski scalar-tensor theory in which  dark matter and dark energy  in the framework of shift-symmetric generalized Galileon theories was described and  an effective unified cosmic fluid was obtained. We calculate the Hubble parameter of the model and we  plot the curve of the  
evolution of the 21-cm temperature in a specific range of the  redshift parameter  $z$ and compare it to  the curve corresponding to $\Lambda$CDM paradigm. 

\end{abstract}

\title{21-cm Brightness Temperature in a Unified-Dark-Sector Horndeski
Theory}

\author{Bertha Cuadros-Melgar$^1$, Thanasis Karakasis$^2$, Eleftherios Papantonopoulos$^2$, Maria Petronikolou$^3$}
\affiliation{$^1$ Escola de Engenharia de Lorena, Universidade de São Paulo, Estrada Municipal do Campinho 100, CEP: 12602-810, Lorena, SP, Brazil \\
$^2$ Physics Department, National Technical University of Athens, 15780 Zografou Campus, Athens, Greece\\
$^3$ Institute for Astronomy, Astrophysics, Space Applications and 
Remote Sensing, National Observatory of Athens, 15236 Penteli, Greece}

\maketitle

\section{Introduction}

Recently the EDGES collaboration observed an absorption profile centered at 78 MHz in the sky averaged radio spectrum \cite{Bowman:2018yin}, whose source is around redshift $z = 17$. As the first stars were formed, they emitted both Lyman-$\alpha$ photons and x-rays. This radiation penetrated the primordial hydrogen gas and changed the excitation state of the 21-cm hyperfine transition, such that photons from the cosmic microwave background (CMB) were absorbed \cite{Pritchard:2011xb}. The predicted signal at frequencies lower than $200 $ MHz  is compatible with the observed one \cite{Bowman:2018yin}. However, the observation indicates a signal with amplitude $0.5 $ K, which is more than a factor of two larger than the largest predictions \cite{Cohen:2016jbh}.

This discrepancy could be addressed either by increasing the background radiation temperature or by cooling the gas during the relevant epoch. The latter possibility has been explored through interactions between baryons and dark matter \cite{Barkana:2018lgd}, which can cool the hydrogen gas and enhance the 21-cm absorption signal. However, Ref. \cite{Munoz:2018pzp} showed that only a small fraction of dark matter can be millicharged while remaining consistent with existing constraints. Related scenarios have also been investigated in Ref. \cite{Widmark:2019cut}, where sub-GeV dark matter with spin-dependent couplings to nucleons or electrons was shown to induce hyperfine transitions and reduce the spin temperature directly, without necessarily lowering the kinetic temperature of the gas.

An interaction in the dark sector is expected to modify the evolution of dark matter and, consequently, the Hubble expansion rate. This, in turn, affects the optical depth of the 21-cm transition. In particular, a smaller value of the Hubble parameter at high redshift would correspond to a larger optical depth and therefore to a stronger 21-cm absorption signal. Thus, even if the baryonic sector follows the standard cosmological evolution, modifications in the dark sector can influence the predicted absorption profile through their effect on $H(z)$ \cite{Bowman:2018yin}.

One of the best probes of the reionization process of Intergalactic Medium (IGM) is the CMB brightness temperature
fluctuations induced by neutral hydrogen through the 21-cm line.
 Therefore, it is important to
examine the effect of the interaction between baryons and dark matter   CMB brightness temperature fluctuations produced by the redshifted hydrogen 21cm line. The hydrogen 21cm line is caused by
a spin flip of the electron in a neutral hydrogen atom. The neutral hydrogen atoms produce the brightness temperature fluctuation of CMB by 21-cm line absorption from and emission into CMB. The amplitude of the brightness temperature fluctuations depends on the hydrogen density, ionization fraction of hydrogen, and hydrogen temperature.  Therefore, observations of the brightness temperature fluctuations at wavelength $21(1 + z)$cm reveal the density fluctuations and the
ionization process at redshift $z$ \cite{Loeb:2003ya, Bharadwaj, Cooray}.

The 21-cm brightness temperature ($T_{21}$),  is determined by the contrast between the spin temperature $(T_s)$ of neutral hydrogen and the background radiation temperature, together with the optical depth of the 21-cm transition. Since the latter depends on the cosmic expansion rate, modifications of the gravitational sector can leave an imprint on the expected 21-cm signal. In this work, we investigate this possibility within scalar-tensor theories, which constitute a broad class of extensions of General Relativity in which gravity is mediated by the metric together with an additional scalar degree of freedom \cite{scal_tens,Brans:1961sx,Horndeski:1974wa}.

Such theories have been extensively studied in cosmological contexts, as they can lead to significant departures from the standard cosmological evolution and provide possible mechanisms for alleviating some of the existing cosmological tensions \cite{Bellini:2014fua, Kobayashi:2011nu, Deffayet:2011gz, CosmoVerseNetwork:2025alb}.  
In particular, non-minimal derivative (kinetic) couplings between the scalar field and the curvature have been shown to give rise to a rich variety of cosmological behaviors, including transitions between distinct de Sitter phases, phantom-divide crossing, and viable inflationary dynamics, without requiring the introduction of a cosmological constant \cite{Saridakis:2010mf,Kobayashi:2019hrl,Karydas:2021wmx}. More broadly, Horndeski theories accommodate a wide class of dark-energy models with a well-defined phenomenology, offering testable predictions for the evolution of the dark-energy equation of state and the growth of cosmic structure \cite{Kase:2018aps}.
Within this framework, scalar-tensor theories can modify both the effective gravitational coupling and the dynamics of the dark-energy sector, leading to departures from the standard expansion history. Such modifications may include a varying effective Newton's constant or an effective dark-energy equation of state in the phantom regime, and have been shown to contribute to the alleviation of the $H_0$ tension in Horndeski and generalized Galileon models \cite{Petronikolou:2023cwu,Petronikolou:2021shp,Banerjee:2022ynv}.

Beyond their cosmological applications, such theories have also been widely explored in
strong-field regimes \cite{Damour:1996ke, Herdeiro:2020wei, Bakopoulos:2022csr, Antoniou:2017acq, Antoniou:2017hxj} and can lead to modified background evolution, non-trivial scalar configurations and characteristic departures from General Relativity. In particular, scalar-tensor models have been shown to admit black hole solutions with scalar hair and to exhibit a rich phenomenology associated with their stability and dynamics \cite{Martinez:1996gn,Martinez:2004nb,mavwin,Kolyvaris:2009pc,Charmousis:2014zaa,Rinaldi:2012vy,Kolyvaris:2013zfa,Babichev:2013cya,Abdalla:2019,Bakopoulos:2023fmv, Bakopoulos:2023sdm, Bakopoulos:2024hah, Bakopoulos:2024ogt, Bakopoulos:2025eps, Bakopoulos:2025byy, Karakasis:2026wes}.

Here, we focus on the cosmological implications of a specific scalar-tensor model in which the dark sector modifies the evolution of the Hubble parameter. Since $H(z)$ enters directly into the optical depth of the 21-cm transition, the modified expansion history leads to a corresponding change in the predicted brightness temperature. We therefore use the 21-cm signal as a probe of the departure of this model from the standard $\Lambda$CDM cosmology. The work is organized as follows. In Section \ref{horn} we review the Horndeski scalar-tensor theory and in subsection \ref{dmdeun} we study a specific model in the Horndeski theory. In Section \ref{appl}  we study the 21-cm effect in the specific model. Finally, in Section \ref{conc} we conclude.

\section{Horndeski scalar-tensor theory}
\label{horn} 

We will apply the generalized Galileon theory to a cosmological framework.
  As it is known, in order to avoid the Ostrogradsky instability \cite{Ostrogradsky:1850fid}, it is required to keep the equations of motion at second order in derivatives and, thus, the most general four-dimensional scalar tensor theories having second order field equations are described by the action,
\begin{equation}
S=\int d^{4}x\sqrt{-g}\,{\cal L}\,,\label{action1}
\end{equation}
 where $g$ is the determinant of the metric $g_{\mu\nu}$ and the Lagrangian reads \cite{Horndeski:1974wa},
\begin{equation}
{\cal L}=\sum_{i=2}^{5}{\cal L}_{i}\,,\label{Lagsum}
\end{equation}
 with
\begin{align}
&{\cal L}_{2} = K(\phi,X)~,\label{eachlag2}\\
&{\cal L}_{3} = -G_{3}(\phi,X)\Box\phi~,\\
&{\cal L}_{4} = G_{4}(\phi,X)\,
R+G_{4,X}\,[(\Box\phi)^{2}-(\nabla_{\mu}\nabla_{\nu}\phi)\,(\nabla^{\mu}
\nabla^{\nu}\phi)]\,,\\
&{\cal L}_{5} = G_{5}(\phi,X)\,
G_{\mu\nu}\,(\nabla^{\mu}\nabla^{\nu}\phi)\,\nonumber\\&\ \ \
\ \ \ \ -\frac{1}{6}\,
G_{5,X}\,[(\Box\phi)^{3}-3(\Box\phi)\,(\nabla_{\mu}\nabla_{\nu}\phi)\,
(\nabla^{\mu}\nabla^{\nu}\phi)+2(\nabla^{\mu}\nabla_{\alpha}\phi)\,(\nabla^
{\alpha}\nabla_{\beta}\phi)\,(\nabla^{\beta}\nabla_{\mu}\phi)]\,,\label{
eachlag5}
\end{align}
where, for simplicity, we have set the gravitational constant to $\kappa\equiv 8\pi G=1$.
The functions $K$ and $G_{i}$ ($i=3,4,5$) depend on the scalar field $\phi$
and its kinetic energy $X=-\partial^{\mu}\phi\,\partial_{\mu}\phi/2$, while $R$ is the Ricci scalar and $G_{\mu\nu}$ is the Einstein tensor.
$G_{i,X}$ and $G_{i,\phi}$ ($i=3,4,5$) respectively correspond to the
partial derivatives of $G_{i}$ with respect to $X$ and $\phi$,
namely $G_{i,X}\equiv\partial G_{i}/\partial X$ and
$G_{i,\phi}\equiv\partial
G_{i}/\partial\phi$.

The action (\ref{action1}) was first found by Horndeski in
\cite{Horndeski:1974wa}, but it was independently rederived in the framework of Galileon theory. We mention here that in the original version of Galileon theory, shift symmetry plays a crucial role and, hence, the two theories do not coincide. Nevertheless, extending Galileon theory to the so-called generalized Galileon theory, i.e., abandoning the shift symmetry, leads to a complete identification with the Horndeski construction.

\subsection{A specific model}
\label{dmdeun}

A specific model along the lines described in the previous Section was presented in \cite{Koutsoumbas:2017fxp}. In this model a unified description of the dark matter and the dark energy sectors, in the framework of shift-symmetric generalized Galileon theories, was studied. Considering a particular combination of terms in
the Horndeski Lagrangian in which  a cosmological constant or a matter sector was not introduced,  an effective unified cosmic fluid was obtained. 
 This model includes the canonical kinetic term along with two non-trivial Galileon terms and the usual non-minimal derivative coupling. The main motivation for such a construction is to avoid  ghost and Laplacian instabilities \cite{DeFelice:2011bh}.

 In particular, the Lagrangian given in (\ref{Lagsum}) was considered in \cite{Koutsoumbas:2017fxp}, with the function choices $K(\phi,X) = \frac{X}{2} -\frac{ \eta}{2} X^{1/2}$ ,  $G_3(\phi,X) = \frac{\lambda_3}{2}X^{-1/2}$
, $G_4(\phi,X) = \frac{1}{2}$ , $G_5(\phi,X) = -\frac{\lambda_5}{2}\phi$, and for convenience units where $8\pi G = c = \hbar = 1$ were chosen. 

The action (\ref{action1}) then becomes,
\be
\label{actionfin}
S= \int d^4 x \sqrt{-g}\left[\frac{R}{2} +\frac{1}{2} \left(X -
 \eta X^{1/2}\right) -
\frac{\lambda_3 X^{-1/2}}{2}\square\phi
+ \frac{\lambda_5}{2} G_{\mu\nu}\nabla^\mu\phi \nabla^\nu\phi\right]~.
\ee
Without loss of generality $\eta$ was chosen such that its value is
$\eta=1$.

Imposing a flat Friedmann-Robertson-Walker (FRW) background metric of the form,
\begin{eqnarray}
ds^{2}=-dt^{2}+a^{2}(t)\delta_{ij}dx^{i}dx^{j},
\label{metric}
\end{eqnarray}
the two Friedmann equations are, respectively,
\be  \label{frt}
3H^2   = \left(\frac{1}{2} + 9\lambda_5 H^2\right) X - \frac{3\lambda_3 H}{\sqrt{2}}~,
\ee
and
\be  \label{frt2}
-(3H^{2}+2\dot{H})=
-\frac{1}{2} \sqrt{X} + X \left(\frac{1}{2} - 3\lambda_5 H^2 - 2\lambda_5 \dot{H}\right)
-2\lambda_
5 H
\dot{X} + \frac{\lambda_3 \dot{X}}{2\sqrt{2}X},
\ee
with $H(t) = \frac{\dot{a}(t)}{a(t)}$ and $X(t) =
\frac{\dot{\phi}^2(t)}{2}$.
Therefore, in an FRW geometry the Einstein equations
(\ref{frt}) and (\ref{frt2}) reduce to
\begin{align}
3H^{2} &= \rho_U \label{frrho}~, \\
-(3H^{2}+2\dot{H}) &= p_U~,
\label{frpre}
\end{align}
with
\begin{eqnarray}
\label{dent}
&&\rho_U\equiv
 \left(\frac{1}{2} + 9\lambda_5 H^2\right) X - \frac{3\lambda_3 H}{\sqrt{2}}~,\\
\label{pret}
&& p_U\equiv
-\frac{1}{2} \sqrt{X} + X \left(\frac{1}{2} - 3\lambda_5 H^2 - 2\lambda_5 \dot{H}\right)
-2\lambda_
5 H
\dot{X} + \frac{\lambda_3 \dot{X}}{2\sqrt{2}X}~.
 \end{eqnarray}
 Hence, the total equation of state parameter of the Universe reads,
 \begin{eqnarray}\label{EoS}
w_U\equiv\frac{p_U}{\rho_U}=
\frac{-\frac{1}{2} \sqrt{X} + X \left(\frac{1}{2} - 3\lambda_5 H^2 - 2\lambda_5
\dot{H}\right)
-2\lambda_5 H \dot{
X} + \frac{\lambda_3 \dot{X}}{2\sqrt{2}X}}{ (\frac{1}{2} + 9\lambda_5 H^2) X -
\frac{3\lambda_3
H}{\sqrt{2}
}}~.
\end{eqnarray}
In addition, the Klein-Gordon equation reads,
\begin{eqnarray}
-3\sqrt{2} H X^{3/2} + 12\sqrt{2}H X^2 \left(\frac{1}{2} + 3\lambda_5 H^2 + 2
\lambda_5 \dot{H}\right) + 3\lambda_3 H \dot{X} & \nonumber \\
+ 2 X \left[-3\lambda_3 \dot{H} + \frac{1}{2} \sqrt{2} \dot{X} + H^2
\left(-9\lambda_3 + 3\sqrt{2}\lambda_5 \dot{X}\right) \right] &= 0\,,
\label{kgt}
\end{eqnarray}
which, using equations \eqref{dent} and \eqref{pret}, can be rewritten in the standard conservation form, namely,
\begin{eqnarray}
\dot{\rho}_U+3H(\rho_U+p_U)  =  0~.
\label{rhoreqde}
\end{eqnarray}

The fact that the considered action (\ref{actionfin}) exhibits the shift symmetry ensures that the scalar field $\phi$ does not appear in the equations of motion but only its derivatives (i.e., $X(t)$ and $\dot{X}(t)$) do so. This allows to use equations \eqref{frt}, \eqref{kgt} and \eqref{dent} in order to  easily eliminate  these derivatives from equation \eqref{pret}, resulting to an expression of $p_U$ in terms of $\rho_U$, namely,

\begin{small}
\begin{align}
p_U(\rho_U) &= \Big\{[3 \lambda_5 f(\rho_U)-2]
\left\{-3 \lambda_3^2 \{4 + \lambda_5
f(\rho_U) [ 9 \lambda_5 f(\rho_U)-28]\} [\sqrt{3}  \lambda_3 +
       g(\rho_U)]
       \right.\nonumber
       \\ &\;\;\;\; \left.+
     f(\rho_U) [
       3 \lambda_5 f(\rho_U)-2]
       \left\{\sqrt{3} \lambda_3 \lambda_5 f(\rho_U) [
          3 \lambda_5 f(\rho_U)-14] -
       2 \{2 + 3 \lambda_5 f(\rho_U)[ \lambda_5 f(\rho_U)-1]\}
g(\rho_U)\right\}\right\}\Big\}^{-
1}\nonumber \\
&\cdot
\Big\{
-36 \sqrt{3}\lambda_3^5 -
  18  \lambda_3^4 \{2 g(\rho_U) + \lambda_5 f(\rho_U) [ \sqrt{3}  \lambda_3 +
g(\rho_U)]\}  -
   \frac{1}{2} f^{3/2}(\rho_U) [2 -
     3 \lambda_5 f(\rho_U)]^2
     \nonumber
  \\
  &
  \left\{-6 \sqrt{6} \lambda_3 \lambda_5 f(\rho_U) -
     2\sqrt{2}[ \sqrt{3}  \lambda_3 + g(\rho_U)] +
     2 \sqrt{f(\rho_U)} [2  \sqrt{3}  \lambda_3 +
        g(\rho_U)] + \lambda_5 f^{3/2}(\rho_U) [11  \sqrt{3}  \lambda_3 + 2
g(\rho_U)]\right\}
        \nonumber
        \\
        &+
    \frac{3}{2}  \lambda_3^2 f(\rho_U)
  [
     3 \lambda_5 f(\rho_U)-2]
     \left\{10  \sqrt{3}  \lambda_3 +
     6 g(\rho_U) -
     12 \sqrt{2} \lambda_5 \sqrt{
      f(\rho_U)} [ \sqrt{3}  \lambda_3 +
        g(\rho_U)] + \lambda_5 f(\rho_U) [21  \sqrt{3}  \lambda_3 + 19
g(\rho_U)]
\right\}\Big\}~,
\label{prho}
\end{align}
\end{small}
where 
$$f(\rho_U) =\frac{ \rho_U + 6\lambda_5\rho_U^2 +
   \left[6 \lambda_3^2 \rho_U (\frac{1}{2} + 3 \lambda_5
\rho_U)^2\right]^{1/2}}{(\frac{1}{2} +
   3 \lambda_5 \rho_U)^2}\,, \hspace{0.5cm}  \hspace{0.5cm} g(\rho_U) = \left[3 \lambda_3^2 + 2  f(\rho_U)- 3 \lambda_5 f^2(\rho_U)\right]^{1/2}~.$$

Hence, we can now calculate the total equation of state parameter,
\begin{eqnarray}
 w_U(\rho_U)=\frac{p_U(\rho_U)}{\rho_U}
\end{eqnarray}
as a function  of the scale factor and the coupling constants  that appear in the considered action (\ref{actionfin}).  Observing the form of expression $p_U(\rho_U)$ of (\ref{prho}), we can see that there are parameter regions that could give $p_U=0$ for a long time interval, which corresponds to a pressureless component while departing from zero at late times, tending asymptotically to the value $p_U(\rho_U)=-\rho_U$, i.e., to $w_U=-1$.

To obtain a unified description of dark matter and dark energy in the above
shift-symmetric generalized Galileon model, we  solve equation
\eqref{frt} with respect to $X$ obtaining,
\begin{equation}
X = \frac{3 H (\sqrt{2} \lambda_3+2 H)}{2 (\frac{1}{2} +9 \lambda_5 H^2)}~.
\label{Xauxilia}
\end{equation}
Then, substituting the above expression into the Klein-Gordon equation  \eqref{kgt} we get a simple differential equation for $H(t)$, namely,

{\small{
\begin{eqnarray}\label{eq1}
\dot H &=& \frac{\sqrt{3}H^{3/2}\left(\sqrt{2}\lambda_3+2H\right)(1+18\lambda_5 H^2)}{(1+6\lambda_5H^2)(1-54\lambda_5H^2)\lambda_3^2+6\sqrt{2}\lambda_3H(1+10\lambda_5 H^2+144\lambda_5^2H^4)+8H^2(1+18\lambda_5 H^2+216\lambda_5^2 H^4)} \nonumber \\
&& \times \left[\sqrt{(1+18\lambda_5 H^2)(\sqrt{2}\lambda_3+2H)} +6\sqrt{3}\lambda_5H^{5/2}\left(\sqrt{2}\lambda_3-4H\right)-\sqrt{3H} \left(\sqrt{2}\lambda_3+4H\right) \right]\,.
\end{eqnarray}}}

Finally, inserting (\ref{Xauxilia}) and (\ref{eq1}) into equation (\ref{EoS}) we obtain,

{\small{
\begin{eqnarray}\label{weq1}
w_U&=&\Big\{ 3H\left[(1+6\lambda_5 H^2)(1-54\lambda_5 H^2)\lambda_3^2+6\sqrt{2}\lambda_3H(1+10\lambda_5 H^2+144\lambda_5^2 H^4)+8H^2(1+18\lambda_5 H^2+216\lambda_5^2 H^4) \right]\Big\}^{-1} \nonumber \\
&&\times \Big\{ 9H\lambda_3^2[1+32\lambda_5 H^2-36\lambda_5^2 H^4] 
-2\sqrt{6}\lambda_3\left[216\sqrt{3}\lambda_5^2 H^6-90\sqrt{3}\lambda_5 H^4 -3\sqrt{3}H^2 \right. \nonumber \\
&&\quad\left. +18\lambda_5 H^{5/2}\sqrt{(\sqrt{2}\lambda_3+2H)(1+18\lambda_5 H^2)}+H^{1/2}\sqrt{(\sqrt{2}\lambda_3+2H)(1+18\lambda_5 H^2)}\right] +4\sqrt{3}H\left[60\sqrt{3}\lambda_5 H^4 \right. \nonumber \\
&&\quad \left. -18\lambda_5 H^{5/2}\sqrt{(\sqrt{2}\lambda_3+2H)(1+18\lambda_5 H^2)}+2\sqrt{3}H^2-H^{1/2}\sqrt{(\sqrt{2}\lambda_3+2H)(1+18\lambda_5 H^2)}\right] \Big\}
\end{eqnarray}}}
 Hence, as long as we solve the differential equation  (\ref{eq1}), we have the solution for the equation of state $w_U$ from equation (\ref{weq1}).

\begin{figure}[ht]
\includegraphics[width=0.37 \linewidth]{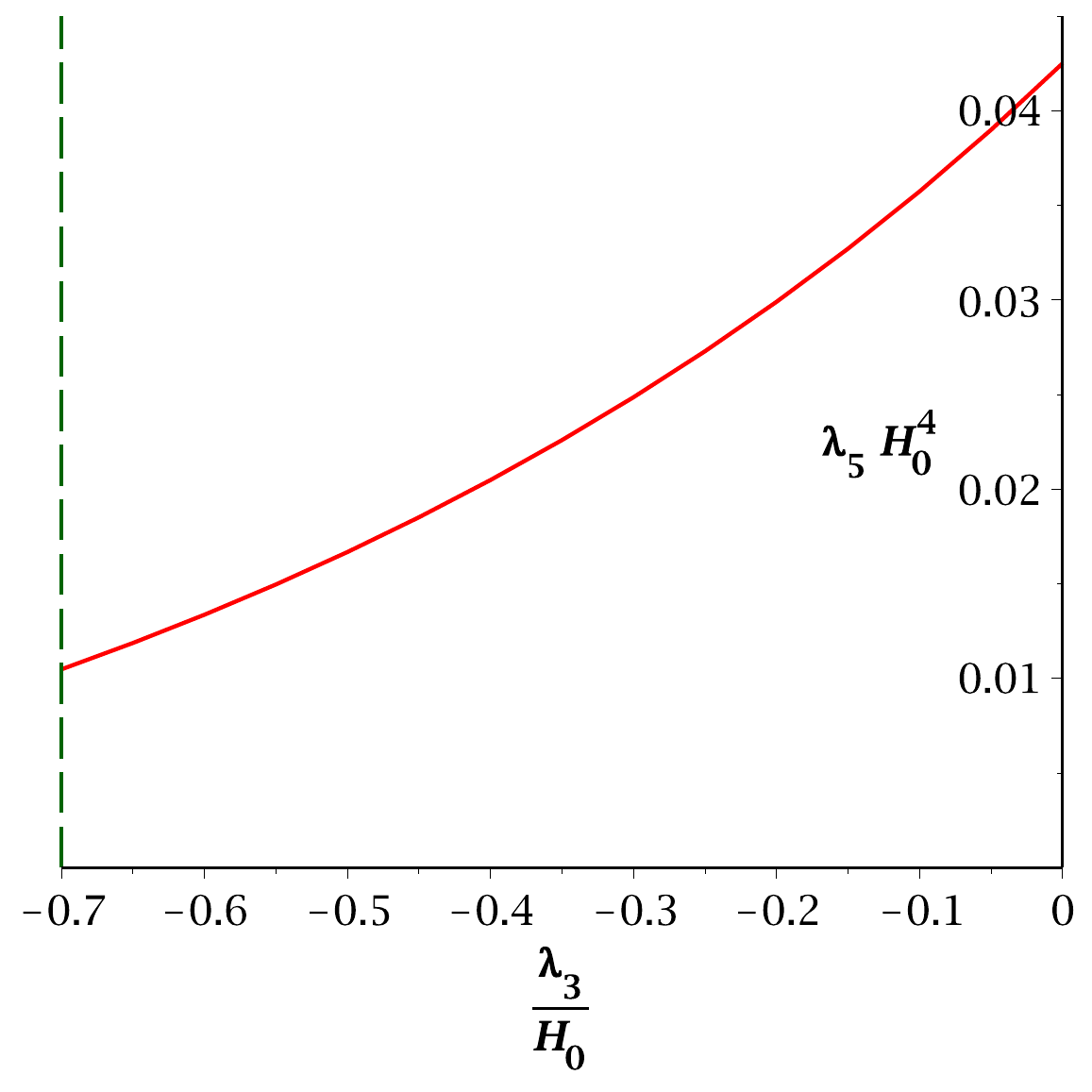}
\caption{\label{couplings}
The one-dimensional parameter subspace of the $(\lambda_3-\lambda_5)$ plane that satisfies the two phenomenological requirements, namely  $H(z=0)\equiv H_0\approx 6 \times 10^{-61}$ (in units where $8\pi G = c = \hbar = 1$) and $w_U(z=0)\approx-0.7$. The dashed vertical line on the left marks the bound in order to avoid Laplacian and ghosts instabilities.} 
\end{figure}

\section{Application to the 21-cm effect}
\label{appl}

Generally, the brightness temperature of the observed 21-cm signal is related to the Hubble parameter $H(z)$, the ratio between the temperature of the background radiation $T_\gamma = 2.73\,(1+z)$, and the spin temperature of the gas $T_s$ as~\cite{Munoz,Mukherjee},
\begin{align}
T_{21} =  \frac{\mathcal{K}}{H(z)(1+z)} \Big{(} 1-\frac{T_\gamma (z)}{T_s (z)} \Big{)}~,\label{21cm}
\end{align}
where, 
\begin{equation}
\mathcal{K} = \frac{3}{32\pi} T_* \, \lambda_{21}^3 A_{10}\, n_{HI} \,,    
\end{equation}
being $T_* =0.068 K $ the energy corresponding to the 21-cm transition, $\lambda_{21}=0.21\,m$, $A_{10}=2.85 \times 10^{-15}s^{-1}$ the Einstein A-coefficient, and $n_{HI}=n_{HI_0} (1+z)^3 \,\bar x_{HI}$ the neutral hydrogen density, whose value at $z=0$, assuming purely cosmic expansion, is $n_{HI_0}=0.19\,m^{-3}$. Also, in the range we are interested in, i.e., $14<z<30$, the fraction of neutral hydrogen is $\bar x_{HI}\approx 1$~\cite{Mukherjee}.
Therefore, if we know $H(z)$, we can calculate the brightness temperature. In our model Eq.(\ref{eq1}) cannot be analytically solved in general. Therefore,  we have to proceed to numerical elaboration in order to extract the solution of $H(t)$.  For convenience, and in order to compare our results with the observational data, we will use the redshift $z=-1+a_0/a$ as the independent variable, setting the current scale factor  $a_0$ to 1 (thus, $\dot{H}
= -(1+z)H(z)H'(z)$, with prime denoting derivatives with respect to $z$).

We set the present value (i.e., at $z=0$) of $H$ to $H_0=\frac{h}{3000} Mpc^{-1}$, with the dimensionless constant $h$ being around $0.69$, which in units where
$8\pi G = c = \hbar = 1$ used above reads  as  $H_0\approx 6 \times 10^{-61}$.
Additionally, we set the present value of the total equation of state of the Universe to $w_U(z=0)\approx-0.7$, according to observations \cite{Ade:2015xua}. Hence,  the above two phenomenological requirements are satisfied by an infinite number of pairs of the remaining parameters $\lambda_3$ and $\lambda_5$, lying in a curve of the $(\lambda_3-\lambda_5)$ plane, shown in Fig.~\ref{couplings}. The three representative values of $\lambda_3$ were chosen within the stable region identified in Fig.~\ref{couplings}, spanning a sufficiently wide range to illustrate the sensitivity of the predicted $T_{21}$ signal to the coupling strength while remaining consistent with the phenomenological requirements $H(z=0)\approx H_0$ and $w_U(z=0)\approx-0.7$.

\begin{figure}[!]
\begin{center}
\includegraphics[width=0.50 \linewidth]{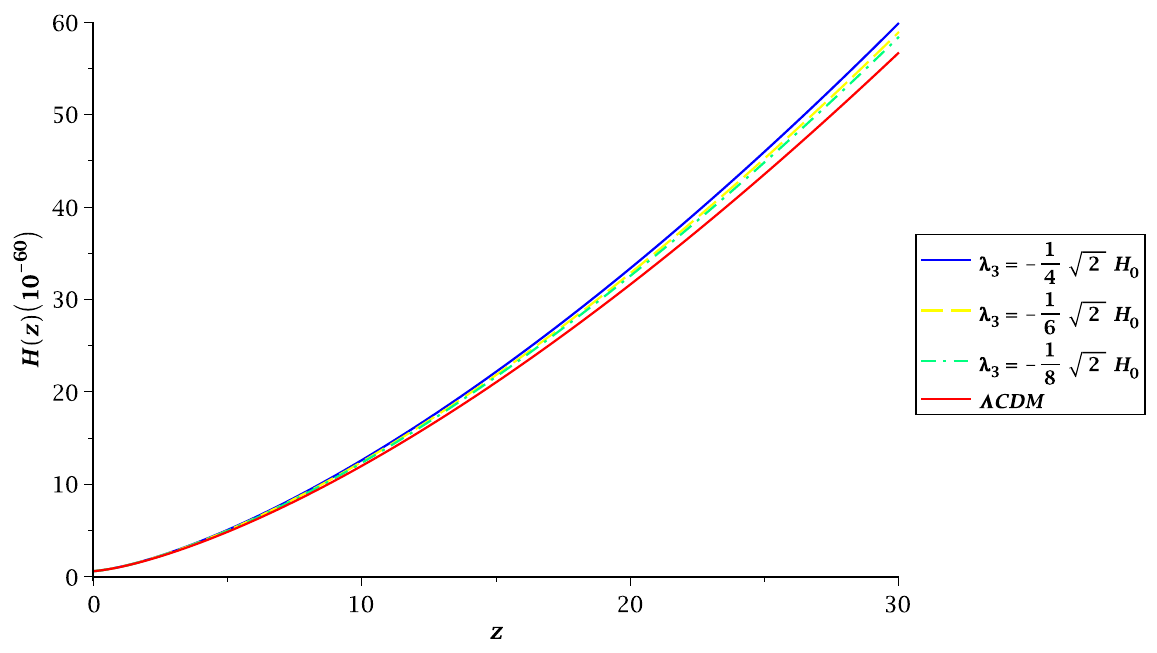}
\includegraphics[width=0.43 \linewidth]{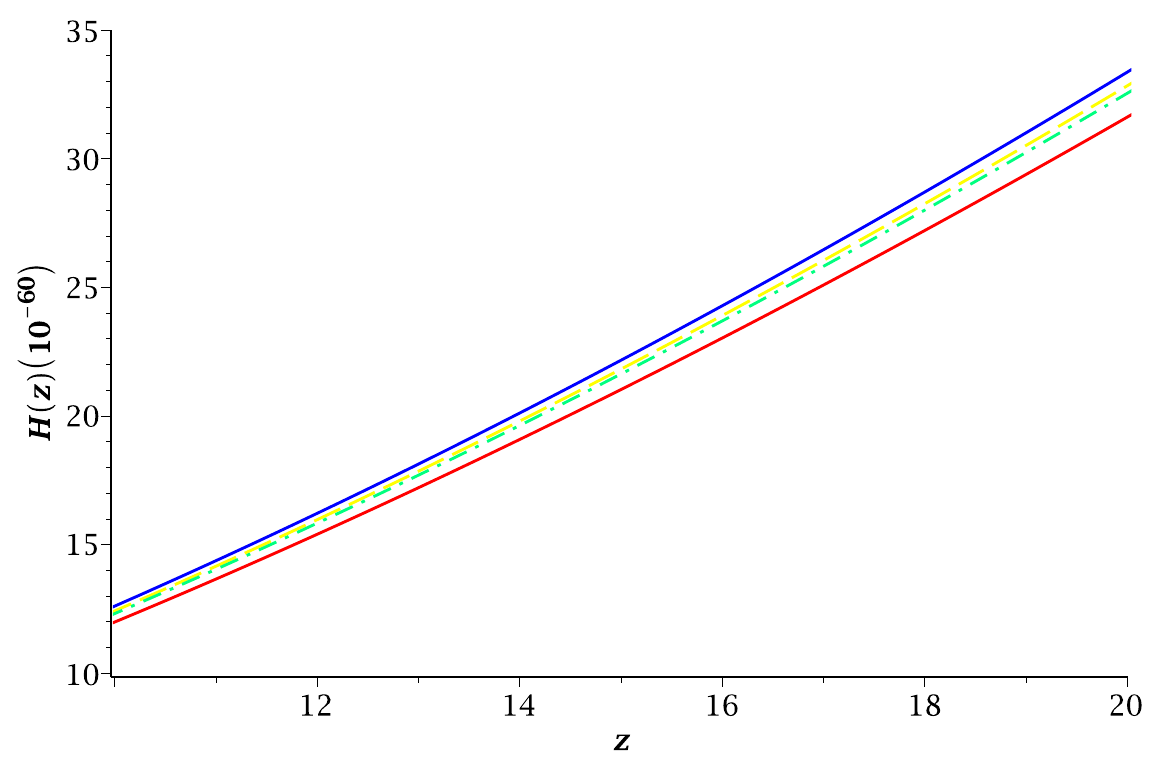}
\caption{\it Left graph: Evolution of the Hubble parameter as a function of the redshift $z$ for the model \eqref{actionfin}, for various values of $\lambda_3$ in units where $8\pi G = c = \hbar = 1$. The present value of the
total equation of state of the universe is set to $w_U(z=0)\approx-0.7$ and the present value of $H$ to $H_0\approx 6 \times 10^{-61}$, which determines accordingly the value of $\lambda_5$ (see Fig.~\ref{couplings}) used here. In the graph the corresponding curve of $\Lambda$CDM paradigm was added. Right graph: Zoom of the region $10<z<20$.} \label{results}
\end{center}
\end{figure}

\begin{figure}[ht!]
\begin{center}
\includegraphics[width=0.5 \linewidth]{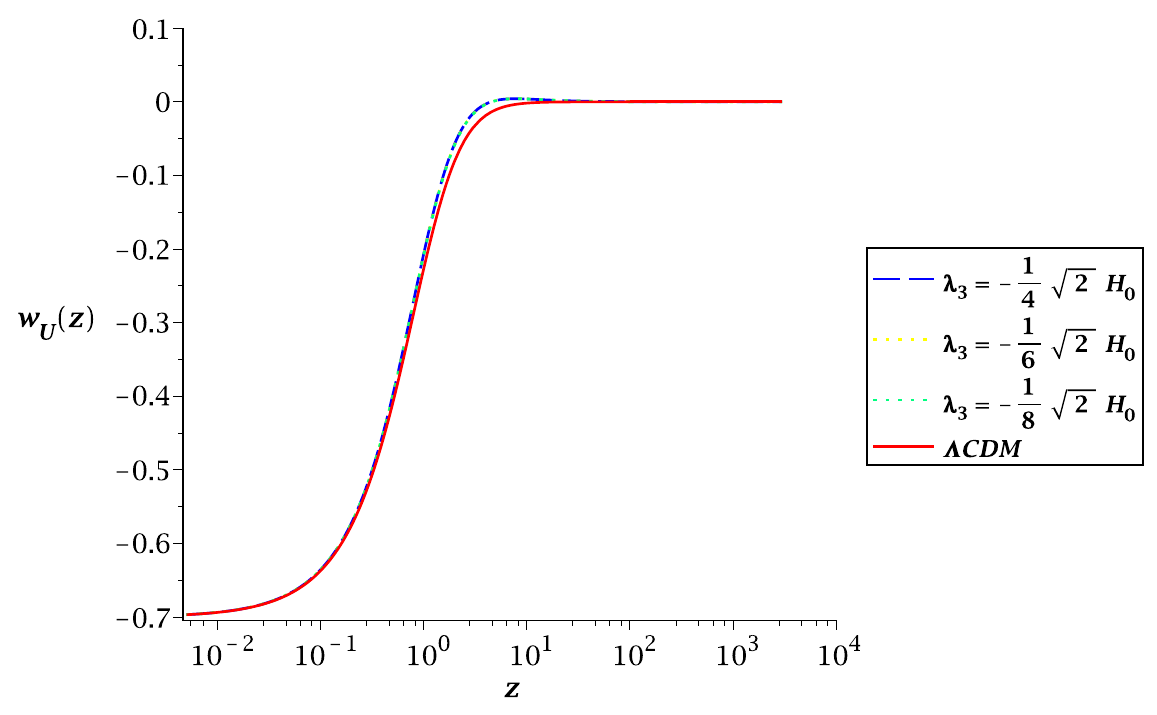} 
\includegraphics[width=0.4 \linewidth]{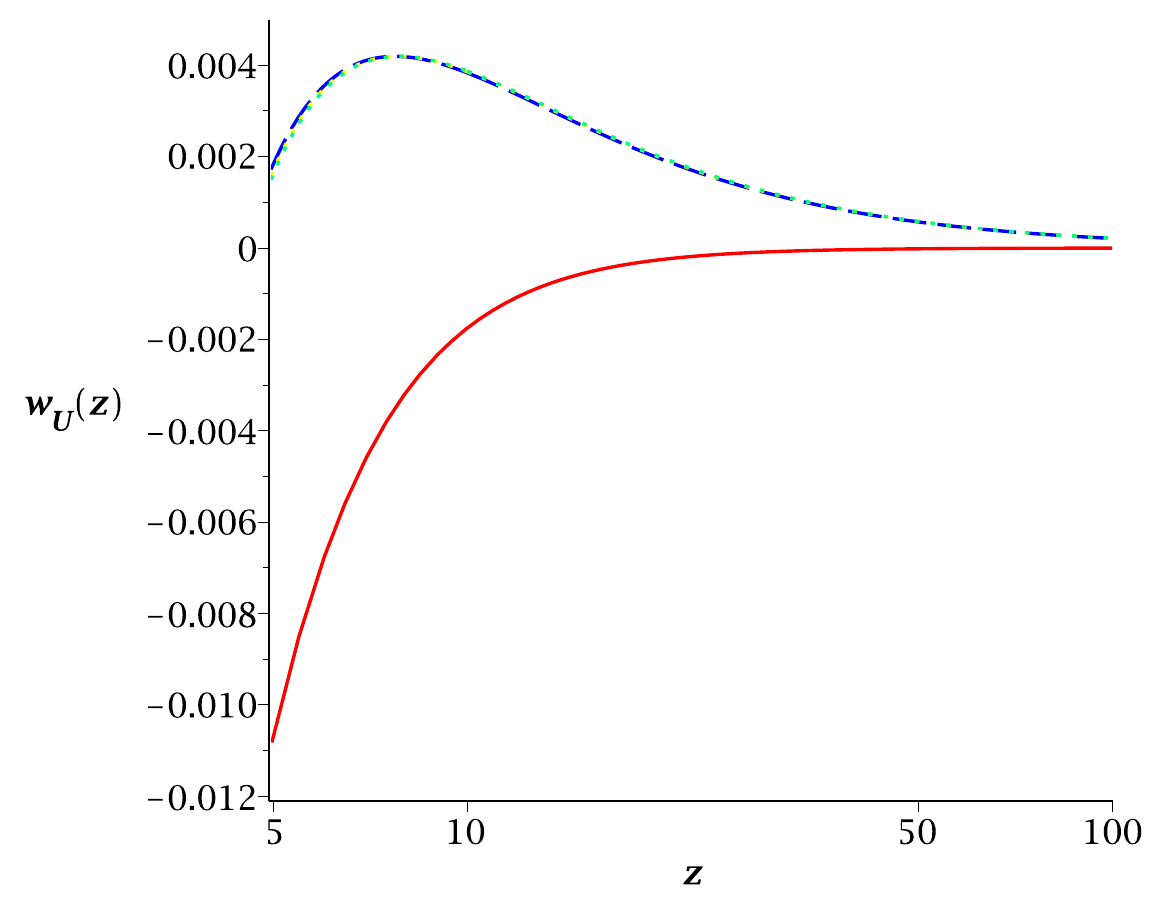} 
\caption{\it  Evolution of $w_U(z)$  in logarithmic scale up to
$z\sim3000$ (left) and a zoom of the region $5<z<100$ where the Horndeski models can be clearly distinguished from $\Lambda$CDM paradigm. } \label{logplot}
\end{center}
\end{figure}

In Fig. \ref{results} we show the solution for $H(z)$ from Eq. 
(\ref{eq1}) for three choices of $\lambda_3$. The corresponding values of $\lambda_5$ were extracted from the existence curve in Fig. \ref{couplings}. For comparison, we additionally depict the corresponding evolution for the $\Lambda$CDM model. We notice that the Horndeski model yields practically the same Hubble parameter at very low redshifts, but begins to diverge as $z$ grows, as shown in the zoom of the region of interest. 
Moreover, because the derivative term acts like a friction term, it was found that the dark matter is represented by a pressureless fluid with the equation of state parameter $w_U(z)$ remaining zero for a long range of the redshift $z$, as  shown in Fig.~\ref{logplot}. We can also see that the main difference among the models occurs around the region $5<z<30$, as displayed in the zoom. 

\begin{figure}[ht!]
\begin{center}
\includegraphics[width=0.4 \linewidth]{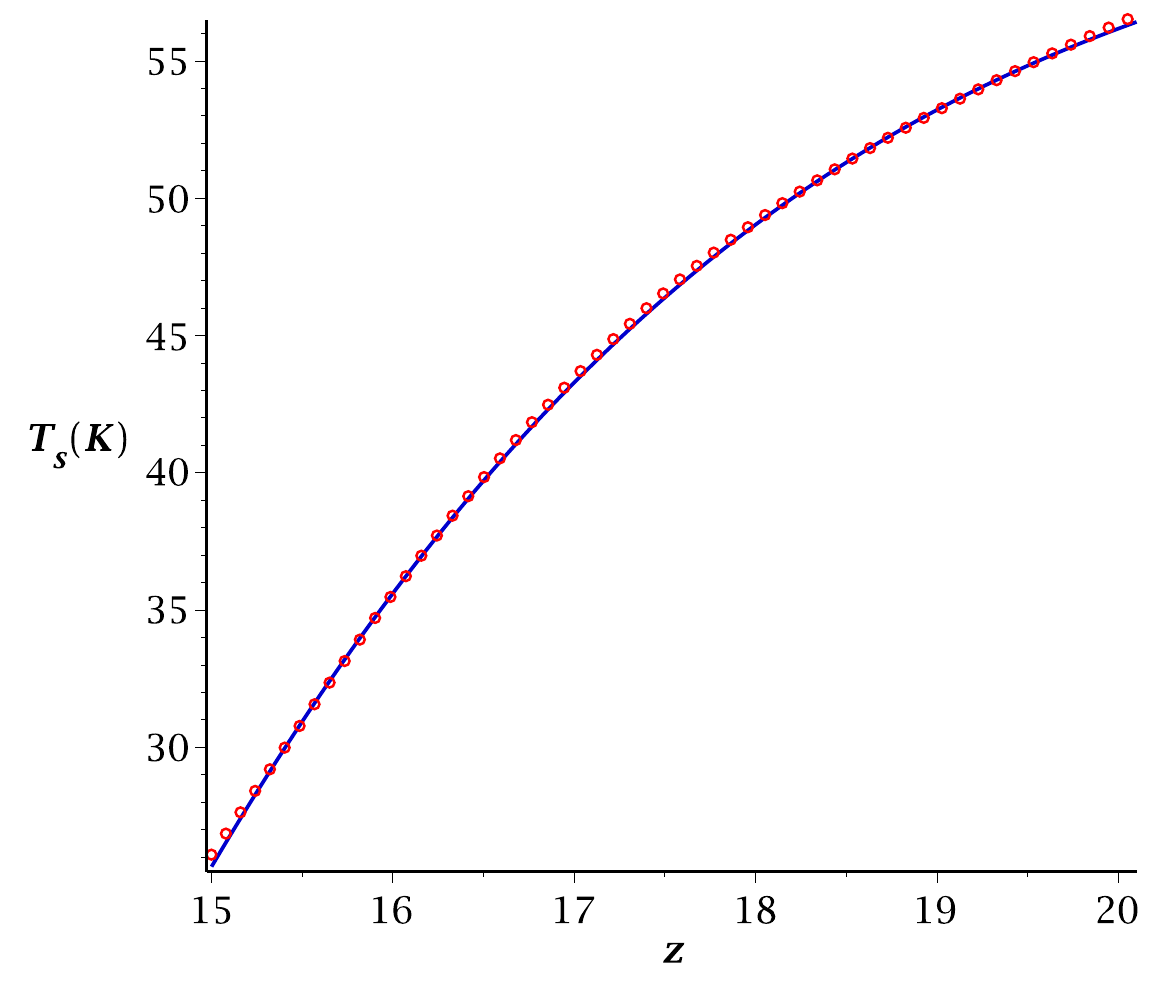}
\caption{\it Spin temperature as a function of the redshift $z$ in the interval $15<z<20$. The blue fit line corresponds to a cubic equation with correlation factor $R^2=0.99987$. } \label{Tspin}
\end{center}
\end{figure}

The spin temperature $T_s$ was obtained using the 21cmFAST software~\cite{Mesinger2011,Mesinger2013}, as shown in Fig.~\ref{Tspin}. Here we show the relevant redshift range ($15<z<20$) and a fit of these data given by a third order polynomial in the redshift $z$, $T_s=az^3+bz^2+cz+d$, where $a=0.05893004719074472$, $b=-3.968576834786584$, $c=90.49451815105779$, $d=-637.7198918788291$, and the correlation factor is $R^2=0.999871356000939$. We should stress that this temperature is an approximation for the modified-gravity model we are using, since 21cmFAST employs the $\Lambda$CDM paradigm in the calculations. This approximation is adopted to isolate the direct impact of the modified expansion history on the 21-cm signal through $H(z)$, while keeping the astrophysical sector fixed. Since the deviations from $\Lambda$CDM in the considered redshift range are relatively small, the $\Lambda$CDM based spin temperature provides a reasonable first-order estimate, although a fully self-consistent treatment would require recalculating $T_s$ within our shift-symmetric Horndeski class background.

\begin{figure}[ht!]
\begin{center}
\includegraphics[width=0.54 \linewidth]{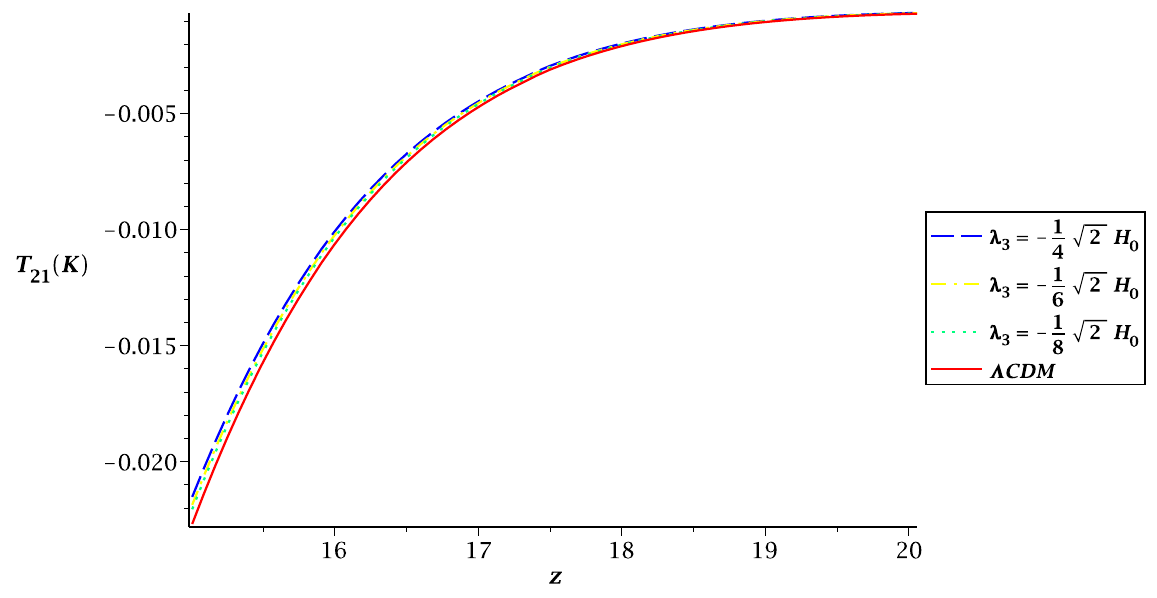}
\includegraphics[width=0.44 \linewidth]{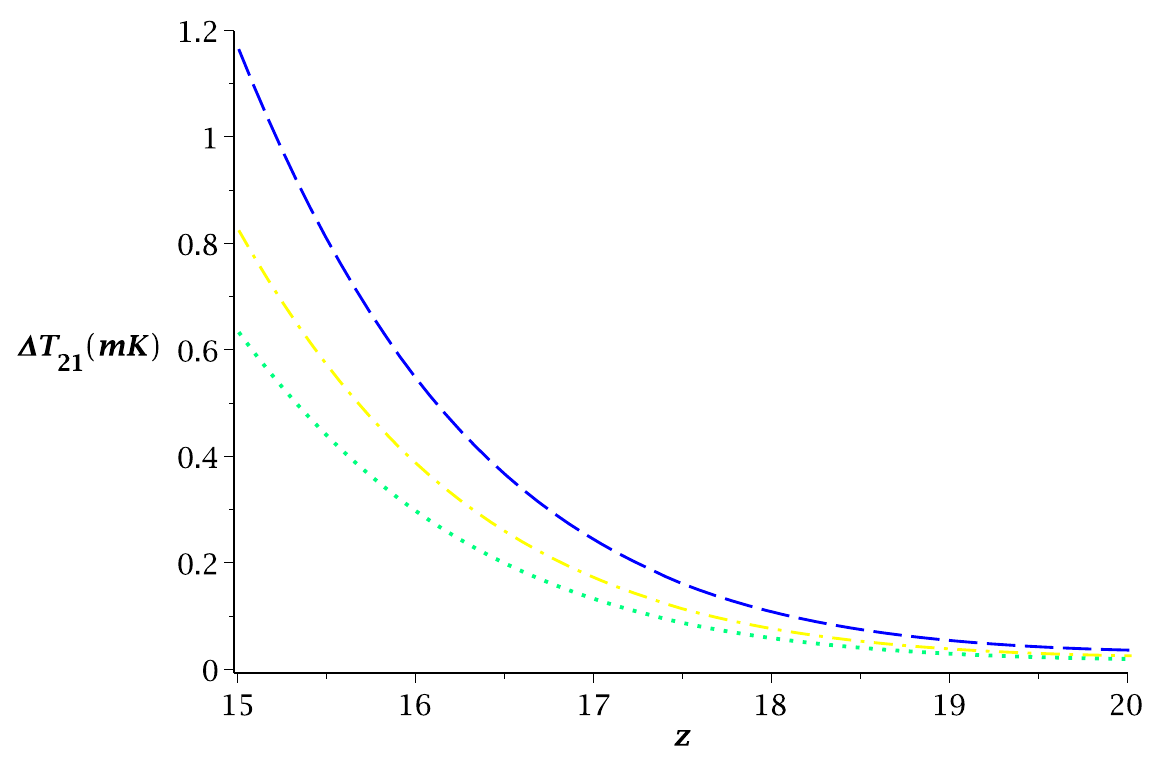}
\caption{\it  Evolution of the 21-cm temperature (left) and deviation from the $\Lambda$CDM model (right) in the range $15<z<20$ for different values of $\lambda_3$. We also included the curve corresponding to $\Lambda$CDM paradigm in the former plot. } \label{T21}
\end{center}
\end{figure}

Putting all this information together, we can finally plot the 21-cm temperature from Eq.\eqref{21cm} in Fig.\ref{T21}. As expected, since the spin temperature $T_s$ is lower than the CMB temperature $T_{\gamma}$, the brightness temperature $T_{21}$ is negative, corresponding to an absorption feature. Moreover, our results show that the modified-gravity model produces less absorption than the $\Lambda$CDM model. Also, as $|\lambda_3|$ increases, the absorption decreases and the deviation from $\Lambda$CDM paradigm is more evident, as we can see in $\Delta T_{21}(z)$ plot in Fig. \ref{T21}. 

\section{Conclusions}
\label{conc}

In this work we investigated the impact of a scalar-tensor theory on 21cm brightness temperature. We considered a specific model in Horndeski theory in which  dark matter and dark energy  in the framework of shift-symmetric generalized Galileon theories were described and  an effective unified cosmic fluid was obtained. Since the evolution of the Hubble parameter directly affects the 21-cm brightness temperature, determining H(z) is a central part of our analysis. As the corresponding evolution equation cannot be solved analytically, we obtained the Hubble parameter numerically.

Using the numerically determined Hubble parameter of the model, we evaluated the evolution of the 21-cm brightness temperature over the redshift interval $15<z<20$ and compared the result with the corresponding $\Lambda$CDM prediction. The brightness temperature depends on the contrast between the spin temperature of neutral hydrogen and the background radiation temperature, as well as on the cosmological expansion rate through the optical depth. The spin temperature was obtained from 21cmFAST and used as a first approximation for the modified-gravity scenario, while the effect of the Horndeski model was introduced through the modified Hubble parameter. The resulting evolution of $T_{21}$, shown in Fig. \ref{T21}, indicates that the Horndeski model predicts a slightly weaker absorption signal than $\Lambda$CDM over the considered redshift range. 

The significance of this analysis lies in showing that the 21-cm signal can be sensitive to modifications of the cosmological expansion history induced by alternative theories of gravity. Even though the deviation from the $\Lambda$CDM prediction is relatively small in the present model, the result demonstrates that the 21-cm brightness temperature can provide an additional cosmological observable for testing scalar-tensor theories during a redshift regime that is otherwise difficult to probe. In this sense, our calculation provides a first quantitative estimate of the direct effect of the modified Hubble expansion on the 21cm signal and establishes a basis for future, fully self-consistent studies in which both the background evolution and the thermal history of the gas are computed within the modified gravity framework.

\section*{Acknowledgments}

B. C.-M. thanks A. Marins and L. H. González-Villegas for enlightening
discussions. The authors acknowledge the contribution of 
the COST Action CA21136 “Addressing observational
tensions in cosmology with systematics and fundamental
physics (CosmoVerse)”.

\end{document}